\documentclass[11pt,a4wide]{article}

\usepackage{amsmath}
\usepackage{amsfonts}
\usepackage{amsbsy}
\usepackage{hyperref}
\usepackage{epsfig}
\usepackage{latexsym}
\usepackage{color,ulem}
\input amssym.def
\input amssym.tex

\usepackage{graphicx} 
\usepackage{tikz}
\usepackage{adjustbox}

\advance \topmargin by -\headheight
\advance \topmargin by -\headsep
\evensidemargin \oddsidemargin
\advance\hoffset by -3mm  
\advance\voffset by  -8mm  
\def\bbbc{{\mathchoice {\setbox0=\hbox{$\displaystyle\rm C$}\hbox{\hbox
to0pt{\kern0.4\wd0\vrule height0.9\ht0\hss}\box0}}
{\setbox0=\hbox{$\textstyle\rm C$}\hbox{\hbox
to0pt{\kern0.4\wd0\vrule height0.9\ht0\hss}\box0}}
{\setbox0=\hbox{$\scriptstyle\rm C$}\hbox{\hbox
to0pt{\kern0.4\wd0\vrule height0.9\ht0\hss}\box0}}
{\setbox0=\hbox{$\scriptscriptstyle\rm C$}\hbox{\hbox
to0pt{\kern0.4\wd0\vrule height0.9\ht0\hss}\box0}}}}

\makeatletter
\@addtoreset{equation}{section}
\makeatother

\begin{document}

\hfuzz=100pt \title{{\Large \bf{{On the Tidal and Frame-Drag Fields in General Relativity}}}}
\author{
Rahulkumar Solanki \footnote{E-mail: rahulscify@gmail.com}}
\date{\today}
\maketitle

\begin{abstract}
\noindent
It is known that in vacuum, two infinitesimally separated timelike observers under free fall notice a relative acceleration due to the tidal field (or the `electric' part of the Weyl tensor), and if they are carrying inertial guidance gyroscopes, they notice relative precession between these gyroscopes due to the frame-drag field (or the `magnetic' part of the Weyl tensor). The frame-drag field thus has no Newtonian analog. In this paper, for a family of timelike observers carrying inertial gyroscopes and forming a congruence, we define these fields as constraint equations in the presence of matter using (1+3) covariant splitting of spacetime. Unlike Newtonian gravity, the relative velocities of the observers around a closed contour formed by their spatial separations (and a temporal vector that closes this contour if the congruence is not hypersurface orthogonal) do not add up to zero due to the frame-drag field and momentum density of the fluid. Moreover, the relative orientations of the gyroscope spins around the same contour do not generally cancel out due to the tidal field, the fluid variables (anisotropic pressure and energy density), and the spatial cross-product of relative velocities associated with the separation vectors that form the contour. Additionally, it is shown that even in the absence of the frame-drag field, neighboring observers see relative precession between their gyroscopes when their relative velocity and acceleration are non-parallel due to differential Thomas precession.
\end{abstract}

\section{Introduction}
The change in a vector $A^{\mu}$ under parallel transport around a closed infinitesimal contour is given by the Riemann tensor 
\begin{equation}
	\label{Curvature}
	\Delta A^{\mu}=-\frac{1}{2}\:R^{\mu}_{\;\;\nu\rho\sigma}\:A^{\nu}\Delta f^{\rho\sigma},
\end{equation}
where $\Delta f^{\rho\sigma}$ is the area element bounded by the contour \cite{Landau:1982dva,Misner:1973prb}. Here, Greek indices represent spacetime components and Roman indices represent spatial components. 
Since the length of a vector does not change under parallel transport, $A_\mu\:\Delta A^\mu$ is zero.
The Riemann tensor can also be defined using the Ricci identity, 
\begin{equation}
    \label{Ricci_identity}
    R^{\mu}_{\;\;\nu\rho\sigma}\:u^{\nu}=\left(\nabla_{\rho}\nabla_{\sigma}-\nabla_{\sigma}\nabla_{\rho}\right)u^{\mu},    
\end{equation}
which can be generalized to tensor fields 
\begin{equation}
    \label{Ricci_identity_II}
    \left(\nabla_{\rho}\nabla_{\sigma}-\nabla_{\sigma}\nabla_{\rho}\right)A^{\mu\nu}=R^{\mu}_{\;\;\delta\rho\sigma}\:A^{\delta\nu}+R^{\nu}_{\;\;\delta\rho\sigma}\:A^{\mu\delta}.    
\end{equation}
The trace of the Riemann tensor is related to the matter via the Einstein field equations
\begin{equation}
	\label{EFE}
	R^{\sigma}_{\;\;\mu\sigma\nu}=R_{\mu\nu}=8\pi\left(T_{\mu\nu}-\frac{T}{2}g_{\mu\nu}\right),
\end{equation}
and the trace-free part is given by the Weyl tensor
\begin{equation}
	\label{Weyl}
	C^{\mu\nu}_{\quad\rho\sigma}=R^{\mu\nu}_{\quad\rho\sigma}-16\pi\; \delta^{[\mu}_{\quad[\rho} T^{\nu]}_{\quad\sigma]} + \frac{16\pi}{3}\; T \; \delta^{[\mu}_{\quad[\rho} \delta^{\nu]}_{\quad\sigma]}.
\end{equation}
Here, the round brackets over indices indicate symmetrization and square brackets indicate anti-symmetrization.
Moreover, we choose geometric units ($c=G=1$) throughout the paper.

\subsubsection*{\centering{(1+3) covariant splitting of spacetime}}
The Weyl tensor can be uniquely decomposed with respect to an observer moving with four-velocity $u^{\mu}$ \cite{Ellis:1971pg,Maartens:1997fg}
\begin{equation}
	\label{Weyl_decomposition}
	C_{\alpha\beta}^{\;\;\;\;\:\gamma\delta}=4\left(u_{[\alpha}u^{[\gamma} + h^{\;\;\;[\gamma}_{[\alpha} \right) E^{\;\;\;\delta]}_{\beta]}+2\epsilon_{\alpha\beta\mu} u^{[\gamma} H^{\delta]\mu} + 2 u_{[\alpha} H_{\beta]\mu} \epsilon^{\gamma\delta\mu},  
\end{equation}
where $\epsilon_{\mu\nu\rho}=\epsilon_{\mu\nu\rho\sigma}u^{\sigma}$ is the spatial permutation tensor, $h_{\mu\nu}=g_{\mu\nu}+u_{\mu}u_{\nu}$ is the spatial projection tensor and $u^\mu u_\mu =-1$. Moreover,
\begin{align}
    E_{\mu\nu}&=C_{\mu\rho\nu\sigma}u^{\rho}u^{\sigma} \label{EM_Weyl_I}\\
    H_{\mu\nu}&=\frac{1}{2}\epsilon_{\mu\alpha\beta}C^{\alpha\beta}_{\;\;\;\;\:\nu\gamma}u^{\gamma}=\frac{1}{2}\epsilon_{\nu\alpha\beta}C^{\alpha\beta}_{\;\;\;\;\:\mu\gamma}u^{\gamma}. \label{EM_Weyl_II}
\end{align}
Both $E_{\mu\nu}$ and $H_{\mu\nu}$ are symmetric, trace-free and spatial.
The former has a well-known physical interpretation and is called the tidal field \cite{Misner:1973prb}. 
For example, consider two infinitesimally separated freely falling observers in vacuum, initially at rest with respect to each other and their inertial guidance gyroscopes aligned.
These observers will notice the relative acceleration due to the tidal field. 
As shown independently in \cite{Sachs1960}, \cite{Estabrook:1964zk} and \cite{Nichols:2011pu}, they will also notice relative precessions of the gyroscopes due to $H_{\mu\nu}$.
Since these gyroscopes constitute a local inertial reference frame, their relative precessions represent the differential ``dragging" of the inertial frames.
Therefore, it was called the frame-drag field in \cite{Nichols:2011pu}. 
Since both the tidal and frame-drag fields are symmetric and purely spatial, their eigenvalues are real and their eigenvector fields can be chosen to be orthogonal. 
The effect of the tidal field on a test body in vacuum, for example, is to stretch (squeeze) it along its eigenvector fields for negative (positive) eigenvalues. 
And the effect of the frame-drag field is to ``twist" it counterclockwise (clockwise) around the eigenvector fields for negative (positive) eigenvalues.  

The tidal and frame-drag fields are widely known as `electric' and `magnetic' parts of the Weyl tensor respectively due to a striking resemblance of the Weyl tensor satisfying the Bianchi identity with the Faraday tensor satisfying the Maxwell equations (see \cite{Bel,Misner:1973prb,Trumper,Maartens:1997fg} and references therein).

\subsubsection*{\centering{Tidal and Frame-drag fields in the presence of matter}}
In order to write all relationships in the presence of matter, we start with the unique splitting of the energy-momentum tensor in terms of quantities measured by an observer moving with unit four-velocity $u^\mu$ \cite{Ellis:1971pg}.
\begin{equation}
	\label{EM_decomposition}
	T_{\mu\nu} =\rho\: u_{\mu}u_{\nu}+2 q_{(\mu}u_{\nu)}+P h_{\mu\nu}+\Pi_{\mu\nu}.
\end{equation}
Here $\rho$ is the total energy density, $P$ is the isotropic pressure, $q_\mu$ is the energy flux (or momentum density) and $\Pi_{\mu\nu}$ is the anisotropic pressure (or stress) measured by the observer. 
Furthermore, $q_\mu$ and $\Pi_{\mu\nu}$ are purely spatial and the latter is trace-free.
Thus,
\begin{equation}
	\label{fluid_components}
	\rho=T_{\mu\nu}u^{\mu} u^{\nu}\; ,\quad  q_\mu=-T_{\alpha\beta}u^{\alpha}h^{\beta}_{\mu}\; ,\quad P h_{\mu\nu}+\Pi_{\mu\nu}=T_{\alpha\beta}\:h^{\alpha}_{\mu}\:h^{\beta}_{\nu}.
\end{equation} 
Using eq.(\ref{Weyl}), (\ref{EM_Weyl_I}), (\ref{EM_Weyl_II}) and (\ref{EM_decomposition}), one can construct spatial tensors from the Riemann tensor that reduce to the tidal and frame-drag fields in the absence of matter \cite{Costa}:
\begin{align}
    \mathbb{E}_{\mu\nu}&=\quad R_{\mu\rho\nu\sigma}u^{\rho}u^{\sigma}\;=E_{\mu\nu}+\frac{4\pi}{3}\left({\rho}+3P\right)h_{\mu\nu}-4\pi\;\Pi_{\mu\nu}, \label{E_Riemann}\\
    \mathbb{H}_{\mu\nu}&=\frac{1}{2} \epsilon_{\mu\rho\sigma} R^{\rho\sigma}_{\;\;\;\;\nu\delta} u^{\delta}=H_{\mu\nu}-4\pi\epsilon_{\mu\nu\sigma}q^{\sigma}. \label{H_Riemann}
\end{align}
As can be seen, $\mathbb{E}_{\mu\nu}$ is symmetric but not trace-free in general and $\mathbb{H}_{\mu\nu}$ is trace-free but not symmetric in general. 
Interestingly, $\mathbb{H}_{\mu\nu}$ is symmetric and equal to frame-drag field in the presence of matter as long as the energy flux $q_{\mu}$ vanishes, e.g., comoving observers following timelike worldlines of perfect fluid.

\subsubsection*{\centering{Congruence of timelike curves }}
We consider a congruence of timelike observers carrying inertial guidance gyroscopes.
Therefore, there exist velocity $u^\mu$ and spin $s^\mu$ vector fields such that $s_\mu s^\mu=1$ and $s_\mu u^\mu=0$ everywhere. 
Moreover, $s^\mu$ is Fermi-Walker transported along the worldlines (see Sec. \ref{FW_transport} below).  
The velocity gradient $\nabla_\nu u_\mu$ (or the directional derivative) in the temporal direction yields the four-acceleration, while in the spatial direction, it can be decomposed into expansion, shear and vorticity \cite{Ellis:1971pg, Poisson}: 
\begin{equation}
    \label{congruence}
    B_{\mu\nu}=\nabla_{\nu}u_{\mu} =\Theta_{\mu\nu} +\omega_{\mu\nu}-a_{\mu}u_{\nu},
\end{equation}
where 
\begin{equation}
    \label{congruenceII}
    a^\mu=u^{\nu}\nabla_{\nu} u^{\mu}, \quad \Theta_{\mu\nu}=\frac{1}{3}\theta\:h_{\mu\nu} + \sigma_{\mu\nu}, \quad \text{and} \quad \omega_{\mu\nu}=\epsilon_{\mu\nu\sigma}\omega^{\sigma}.
\end{equation}
Here, $a_\mu$ is the four-acceleration, $\theta$ is the expansion, $\sigma_{\mu\nu}$ is the shear and $\omega_{\mu\nu}=\epsilon_{\mu\nu\sigma}\omega^{\sigma}$ is the vorticity.\footnote{In Ref.\cite{Nichols:2011pu}, the eigenvalues associated with the frame-drag field are called vorticity, and their integral curves associated with the eigenvectors are called vortex lines. In order to avoid potential confusion and to maintain a strict analogy with Newtonian hydrodynamics, we follow Ref.\cite{Ellis:1971pg} in calling $\omega^{\mu}$ as vorticity throughout the paper.} 
The shear and vorticity tensors are purely spatial and trace-free. The former is symmetric, whereas the latter is antisymmetric. 

\subsubsection*{\centering{Overview of main results of the paper}}
If the vorticity $\omega^{\mu}$ is zero, one can construct a purely spatial closed contour. 
The change in $u^{\mu}$ under parallel transport around a closed infinitesimal spatial contour is given by the frame-drag field $H_{\mu\nu}$ and the component of momentum density $q_\mu$ residing in the plane of the contour. 
In other words, $\mathbb{H}_{\mu\nu}$ measures the failure of $u^{\mu}$ to be parallel transported around an infinitesimal spatial loop. 
Now, the change in $u^\mu$ under parallel transport from $x^\alpha$ to $\left(x^\alpha+dx^\alpha\right)$, when subtracted from the ordinary change $du^\mu$, gives the difference between $u^\mu$ and its parallel transported value at the same point $\left(x^\alpha+dx^\alpha\right)$, i.e., the covariant change $Du^\mu$. 
If the spatial loop is formed by separation vector fields connecting four observers, and since $Du^\mu$ along spatial separation gives relative velocity, the change in $u^{\mu}$ under parallel transport around this loop is equivalent to the following: the covariant sum of relative velocities of the observers around the loop, unlike Newton's theory, is non-zero in general and is given by $\mathbb{H}_{\mu\nu}$.
(This covariant sum is evaluated in the opposite direction of the parallel transport).
It can be further stated as follows: the difference between the relative velocity of two non-neighboring observers calculated via two different paths along this loop is given by $\mathbb{H}_{\mu\nu}$.
In other words, $\mathbb{H}_{\mu\nu}$ can be written as the spatial curl of the velocity gradient $\nabla_\nu u_\mu$ and represents the path dependence of relative velocity.
Furthermore, since the vorticity is zero, the constant proper time hypersurface is spatial and the change in spin $s^\mu$ under Fermi-Walker transport around a spatial loop is given by the intrinsic curvature of this hypersurface, which is related to the tidal field $E_{\mu\nu}$, the fluid variables (energy density and anisotropic pressure) and the extrinsic curvature via the Gauss-Codazzi equation.  
Although one can align the local `non-rotating' spatial triads carried by neighboring observers, there is path dependence in aligning the local triads beyond that, which depends on the intrinsic curvature of the hypersurface.
If the worldlines are timelike geodesics and hypersurface orthogonal, then the spatial loop will remain spatial along the geodesics. 

If the worldlines are non-geodesics, then they have ``non-gravitational" acceleration and a spatial contour will not remain spatial along the worldlines unless $a^\mu$ is orthogonal to it.
The spatial projection of this contour, however, is still closed due to hypersurface orthogonality, so the previous definitions of tidal and frame-drag fields as spatial constraints remain the same.
However, definitions in terms of temporal evolution change slightly. 
Now, $\mathbb{E}_{\mu\nu}$ gives the relative ``gravitational" acceleration between two infinitesimally separated observers, that is, the difference between the total and non-gravitational relative accelerations, with different proper times of the observers accounted for.
Furthermore, $\mathbb{H}_{\mu\nu}$ gives the relative precessional ``gravitational" angular velocity since even in the absence of gravity these observers see relative precession of their gyros due to differential Thomas precession when their relative velocity and acceleration are non-parallel.

If the vorticity is non-zero, one cannot construct a purely spatial closed contour (unless $\omega^\mu$ is orthogonal to both the separation vectors forming the contour). In other words, the contour constructed using purely spatial vectors will be open, and the displacement vector that closes this contour will be purely temporal.
If $a_\mu =0$, the Fermi-Walker transport reduces to parallel transport, and both $u^{\mu}$ and $s^\mu$ are parallel transported along worldlines.
Therefore, the temporal ``closer of the contour" does not affect the change in $u^{\mu}$ and $s^\mu$ under parallel transport around the closed contour, and the previous definition of the frame-drag field remains the same.
However, the definition of the tidal field as a constraint changes since $\tau=$constant hypersurface is no longer spatial.
Now, the change in $s^\mu$ under Fermi-Walker transport around the closed contour is given by the tidal field, fluid variables (energy density and anisotropic pressure) and the spatial vector triple product between spin and relative velocities of neighboring observers forming the contour. (This product is evaluated at the same worldpoint where the change in $s^\mu$ is calculated.) 
In other words, the sum of relative orientations of inertial gyros around the closed contour in the vacuum is given by the tidal field if the spin is orthogonal to the relative velocities associated with separation vectors forming the contour. 

If both vorticity and acceleration are non-zero, the definition of tidal field remains the same since the spin is Fermi-Walker transported along the worldline.
However, since $u^\mu$ is not parallel transported along the ``closer," its contribution is non-zero and accounts for different proper times when calculating the relative velocity of non-neighboring observers via two different paths.

\section{Physical setup}
Given a congruence of timelike curves, consider a local coordinate transformation $x^\mu=x^\mu(\tau,y^a)$ where $y^1=\xi$, $y^2=\zeta$, $y^3=\chi$ and $\tau$ is the proper time along the curve.
Here, the coordinates $y^a$ are intrinsic to the spacelike hypersurface $\tau=$constant.
If the worldlines are trajectories of fluid particles, then one is transforming from Eulerian to comoving (or Lagrangian) coordinates. 
Now, define
\begin{equation}
    u^\mu=\frac{\partial x^\mu}{\partial \tau},\quad \xi^\mu=\frac{\partial x^\mu}{\partial \xi},\quad \zeta^\mu=\frac{\partial x^\mu}{\partial \zeta} \quad \text{and} \quad \chi^\mu=\frac{\partial x^\mu}{\partial \chi}.
\end{equation}
Thus, $u^\mu$ is the four-velocity field (i.e., unit tangent to the worldlines) and $\xi^\mu\delta\xi$ is the separation vector that connects two worldlines, one at $\xi=\:$constant and the other at $\xi+\delta\xi=\:$constant, at the same proper time $\tau$. 
Although it is cumbersome to keep the factors of $\delta\xi$ or $\delta\tau$ in calculations, we retain them (without redefining the separation vectors) to track the order of terms.
Since these are coordinate lines, these separation vectors are Lie transported along the worldline
\begin{equation}
	\label{Lie_transport_I}
    u^\mu_{\;\: ;\nu}\;\xi^\nu=\xi^\mu_{\;\: ;\nu}\;u^\nu \quad,\quad u^\mu_{\;\: ;\nu}\;\zeta^\nu=\zeta^\mu_{\;\: ;\nu}\;u^\nu \quad,\quad u^\mu_{\;\: ;\nu}\;\chi^\nu=\chi^\mu_{\;\: ;\nu}\;u^\nu
\end{equation}
and along the coordinate lines on $\tau=$constant hypersurface
\begin{equation}
	\label{Lie_transport_II}
    \xi^\mu_{\;\: ;\nu}\;\zeta^\nu=\zeta^\mu_{\;\: ;\nu}\;\xi^\nu \quad,\quad \xi^\mu_{\;\: ;\nu}\;\chi^\nu=\chi^\mu_{\;\: ;\nu}\;\xi^\nu \quad,\quad \zeta^\mu_{\;\: ;\nu}\;\chi^\nu=\chi^\mu_{\;\: ;\nu}\;\zeta^\nu
\end{equation}
Here, a semicolon represents a covariant derivative and a comma represents a partial derivative. 
The directional derivatives can be written as,
\begin{equation}
	\label{Notations_I}
    u^{\nu}\nabla_{\nu}\equiv \frac{D}{D\tau} \quad,\quad \xi^{\nu}\nabla_{\nu}\equiv \frac{D}{D\xi} \quad,\quad \zeta^{\nu}\nabla_{\nu}\equiv \frac{D}{D\zeta} \quad,\quad \chi^{\nu}\nabla_{\nu}\equiv \frac{D}{D\chi}
\end{equation}
It would be more accurate to use $\partial \tau$ instead of $D\tau$, for example. However, the latter is used throughout the paper to emphasize its non-commutative property.

In general, this comoving frame rotates spatially since $\xi^\mu$, $\zeta^\mu$ and $\chi^\mu$ are Lie transported along the worldlines. 
A local non-rotating reference frame can be constructed, for example, using orthonormal tetrads consisting of one timelike and three spacelike vectors that are Fermi-Walker transported along the worldline (see Sec. \ref{FW_transport} below), which becomes a local inertial frame if the worldlines are geodesics. 
Here, the timelike tetrad that corresponds to the local time axis is the unit four-velocity $u^\mu$, while the spacelike triad can be realized using inertial guidance gyroscopes \cite{Hawking, Misner:1973prb}. The dimensions of an inertial gyroscope are considered ``point-like," meaning that any force applied to it acts on its center of mass, resulting in zero torque. We generically associate the spin vector field $s^\mu$ with the inertial guidance gyroscopes carried by the congruence of timelike observers.

\begin{figure}[!ht]
\centering
	\hspace*{-1cm}
	\includegraphics[scale=0.47]{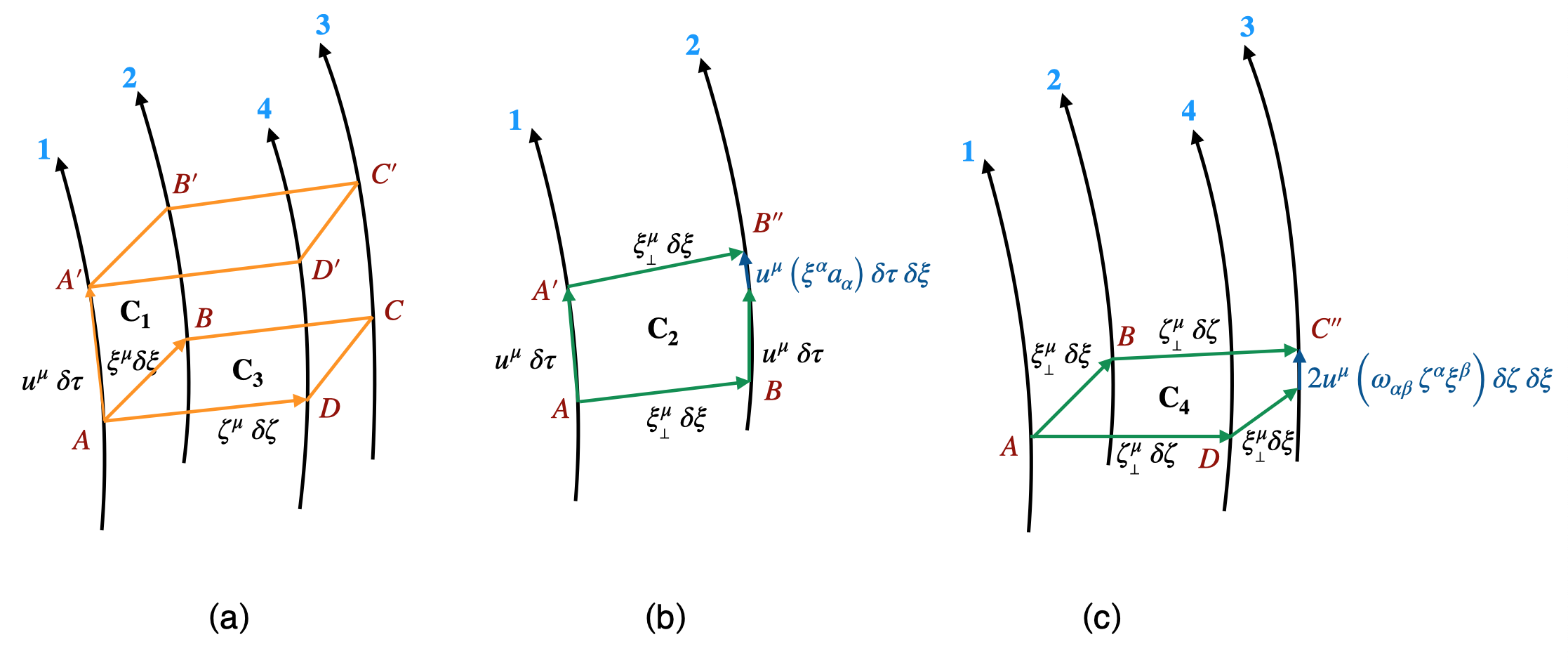}
\caption{\label{Physical_setup_I} The physical setup: Here $(a)$ represents worldlines of four timelike observers. 
This family of timelike curves is parameterized such that $\tau$ is the proper time, the separation vector from the observer $1$ to $2$ is $\xi^{\mu}\:\delta\xi$ and from $1$ to $4$ is $\zeta^{\mu}\:\delta\zeta$. 
The contours \textbf{C}$_\mathbf{1}$, $AA'D'D$ and \textbf{C}$_\mathbf{3}$ are closed, that is, the covariant change in $u^{\mu}$ along $\xi^{\nu}$ (or $\zeta^{\nu}$) is the same as covariant change in $\xi^{\mu}$ (or $\zeta^{\mu}$) along $u^{\nu}$ and the covariant change in $\xi^{\mu}$ along $\zeta^{\nu}$ is equal to the covariant change in $\zeta^{\mu}$ along $\xi^{\nu}$. 
(b) One can choose, for example, $\xi^{\mu}$ and $\zeta^\nu$ to be purely spatial at the worldpoint $A$. However, they will not remain spatial along the worldline (e.g., at the worldpoint $A'$) unless it is a geodesic. In other words, a contour formed by purely spatial and temporal vector fields won't be closed in general. The ``closer of the contour" is given by the Lie bracket \cite{Misner:1973prb}:  $u^\nu \nabla_\nu \xi^{\mu}_{_\perp}-\xi^{\nu}_{_\perp}\nabla_{\nu} u^{\mu}=u^{\mu} (\xi_\nu a^{\nu}).$
(c) A purely spatial contour won't be closed unless the congruence is hypersurface orthogonal. The ``closer of the contour" here is given by the Lie bracket \cite{Misner:1973prb}:   
$\xi^{\nu}_{_\perp} \nabla_{\nu} {\zeta_{_\perp}^{\mu}}-\zeta^{\nu}_{_\perp} \nabla_{\nu} {\xi_{_\perp}^{\mu}}=2u^{\mu} \left(\omega_{\alpha\beta} \zeta^{\alpha} \xi^{\beta}\right)$.
Since the closer of the contour is already a second order term, one can evaluate it at point $A$ rather than point $C''$ (or $B''$). If the wordlines are geodesic but vorticity is non-zero, then 
$u^{\mu}$ is parallel transported along the geodesic so the closer of the contour won't have any contribution in calculating change in $u^\mu$ under parallel transport around \textbf{C}$_\mathbf{4}$.
}
\end{figure}

\subsubsection*{\centering Relative velocity and acceleration}
From Fig.\ref{Physical_setup_I}, the spatial separation between observers $1$ and $2$ is given by
\begin{equation}
	\label{spatial_separation}
	\xi^\mu_{_\perp}\delta\xi=h^{\mu}_{\nu}\:\left(\xi^{\nu}\delta\xi\right).
\end{equation}
Write the covariant change in spatial direction as 
\begin{equation}
	\label{spatial_directional_derivative}
    \xi^{\nu}_{_\perp}\nabla_{\nu}\equiv\frac{D}{D\xi} + \left(\xi^\nu u_\nu\right)\frac{D}{D\tau} \equiv \frac{D}{D\xi_{_\perp}}.
\end{equation}
Now, the relative velocity of the observer $2$ with respect to $1$ can be defined as the spatial rate of change (in terms of proper time) of spatial separation or the covariant change in four-velocity field along the spatial separation \cite{Ellis:1971pg,Ehlers:1961xww} which using eq.(\ref{congruence}), (\ref{spatial_separation}) and (\ref{spatial_directional_derivative}) simplifies to
\begin{equation}
    \label{relative_velocity}
    h^{\mu}_{\nu}\: \frac{D\xi^{\nu}_{_\perp}}{D\tau}\:\delta\xi=\frac{Du^\mu}{D\xi_{_\perp}}\:\delta\xi=\Big(\Theta^{\mu}_{\nu}+\omega^{\mu}_{\;\;\nu}\Big) \xi^{\nu}\:\delta\xi.
\end{equation}
Moreover, the total relative acceleration of observer $2$ with respect to $1$ is spatial rate of change of spatial relative velocity:
\begin{equation}
    \label{relative_acceleration}
    h^{\mu}_{\nu}\: \frac{D}{D\tau} \left(h^{\nu}_{\sigma}\frac{D\xi^{\sigma}_{\perp}}{D\tau}\right)\:\delta\xi=h^{\mu}_{\nu}\: \frac{D}{D\tau} \left(\frac{Du^{\nu}}{D\xi_{\perp}}\right)\:\delta\xi. 
\end{equation}
Since the separation vector $\xi^\mu$ is Lie transported along the curve, even if it is chosen to be spatial initially, it cannot remain spatial in general, i.e.,
\begin{equation}
    \frac{D}{D\tau} (u_\mu \xi^\mu)= a_\mu \xi^\mu,
\end{equation}
does not vanish unless the curve is a geodesic or $\xi^\mu$ is orthogonal to the acceleration.\footnote{An analogy to the geometry of space may be helpful here. Consider a rainbow arch ladder in a playground. The two arches are analogous to worldlines, and steps are analogous to separation vectors. Moreover, the tangent and normal to an arch at any point are analogous to velocity and acceleration, respectively. Here, the steps are orthogonal to both the tangent and the normal, connecting the two arches at the same arc length, so the ``closer of the contour" is zero (see Fig. \ref{Physical_setup_I} (b)).
}

\begin{figure}[!ht]
\centering
\includegraphics[scale=0.35]{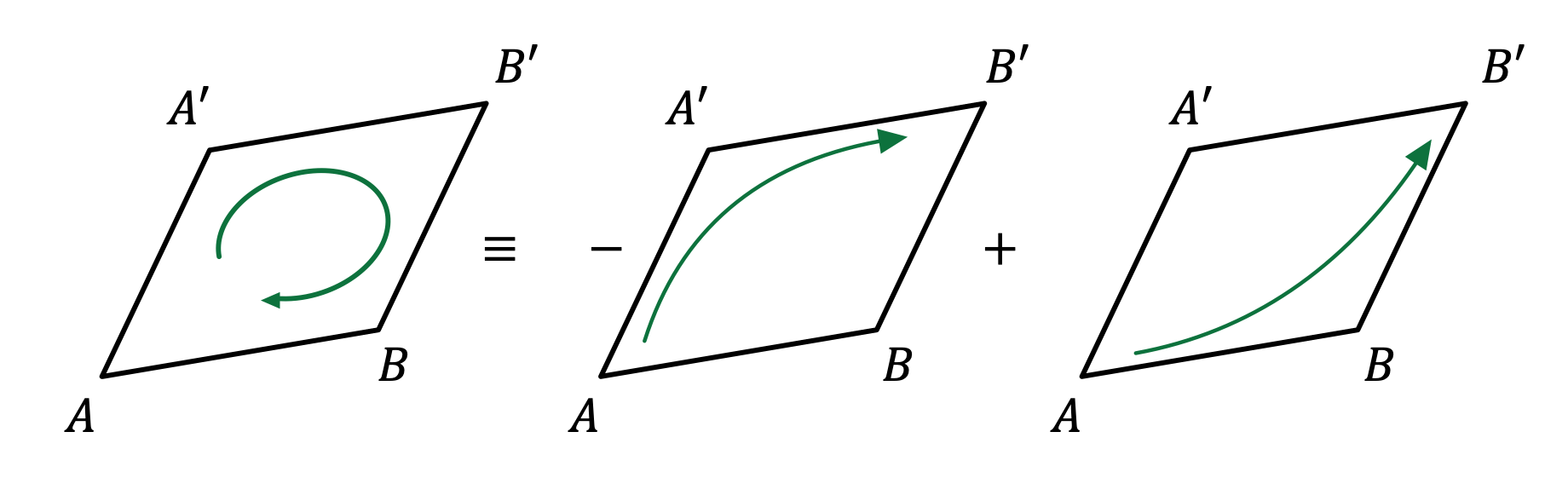}
\caption{\label{Physical_setup_II}To write the relevant formulae pictorially, we show the equivalence between definitions of the Riemann tensor using the Ricci identity, directional derivatives and parallel transport around the loop. The change in $u^{\mu}$ under parallel transport from $x^\sigma$ to $(x^\sigma+dx^\sigma)$ is given by $\delta u^\mu=-\Gamma^{\mu}_{\nu\sigma}\:u^{\nu}dx^\sigma$. 
And the difference between $u^\mu$ at $(x^{\sigma}+dx^{\sigma})$ and its value under parallel transport from $x^\sigma$ to $(x^\sigma+dx^\sigma)$, i.e., the covariant change $Du^\mu$, is given by $(du^\mu-\delta u^\mu)$ \cite{Landau:1982dva}.
Now, the change in $u^{\mu}$ under parallel transport around the closed infinitesimal loop $AA'B'BA$ is $\Delta u^{\mu}=\oint \delta u^{\mu}$. It can be rewritten as, $\Delta u^{\mu}=-\oint Du^\mu=-\oint \nabla_{\nu}u^\mu\;dx^\nu$ since $\oint du^\mu=0$. Since the loop is infinitesimal, applying Stoke's theorem \cite{Landau:1982dva} gives $\Delta u^{\mu}=-\frac{1}{2}(\nabla_{\sigma}\nabla_{\nu}u^\mu-\nabla_{\nu}\nabla_{\sigma}u^\mu)\Delta f^{\sigma\nu}$ which yields the Ricci identity (\ref{Ricci_identity}) when compared with (\ref{Curvature}). 
Additionally, $\Delta u^\mu$ can also be found by calculating covariant difference of $u^\mu$ at $B'$ along two separate paths, i.e., $\Delta u^{\mu}=-\int_{c_1} Du^{\mu}+\int_{c_2} Du^{\mu}$ where $c_1$ is path $AA'B'$ and $c_2$ is $ABB'$.}
\end{figure}

\section{Fermi-Walker transport around a closed contour}
\label{FW_transport}
An inertial guidance gyroscope carried by a timelike observer can be modeled by assigning to it a unit spin vector $s^\mu$ that is spatial and `non-rotating' along the worldline.
Although the orthogonality between two vector fields is preserved under parallel transport, $s^\mu$ cannot remain spatial if parallel transported along the worldline since $u^\mu$ is not parallel transported. 
In other words, both $s^\mu$ and $u^\mu$ follow a different transport rule, known as Fermi-Walker transport, and it reduces to parallel transport if the curve is a geodesic. 
Under this transport, a spatial (temporal) vector remains spatial (temporal) and there are no spatial rotations. 
A vector field $A^\mu$ is Fermi-Walker transported if its Fermi-Walker (or Fermi) derivative \cite{Misner:1973prb, Hawking}
\begin{equation}
    \label{Fermi_derivative}
       \frac{D_{_F} A^{\mu}}{D\tau}  = \frac{DA^{\mu}}{D\tau}+\left(a^{\mu}u^\nu-a^{\nu}u^{\mu}\right)A_\nu,
\end{equation}
vanishes.\footnote{Assume that $A^\mu$ is a unit spacelike vector field along the timelike curve, not necessarily spatial. 
Since $u^\mu$ is Fermi-Walker transported along the timelike curve and the Fermi derivative reduces to ordinary derivative when it acts on a scalar field,
\begin{equation}
    \label{Fermi I}
    \frac{d}{d\tau} (u_\mu A^\mu)=u_\mu\frac{DA^\mu}{D\tau}+a_\mu A^\mu =u_\mu\frac{D_{_F}A^\mu}{D\tau}\quad \implies \quad\frac{D_{_F}A^\mu}{D\tau}=\frac{DA^\mu}{D\tau}-u^\mu (A_\nu a^\nu)+\text{spatial term}.
\end{equation}
Since the Fermi derivative reduces to covariant derivative when $a^\mu=0$, the spatial term must be related to $a^\mu$.
Furthermore, since $A^\mu A_\mu=1$,
\begin{equation}
    \label{Fermi II}
    \frac{d}{d\tau} (A^\mu A_\mu)=2A_\mu \frac{DA^\mu}{D\tau}=2A_\mu \frac{D_{_F}A^\mu}{D\tau}=0.
\end{equation}
From the eq. (\ref{Fermi I}) and (\ref{Fermi II}) one finds,
\begin{equation}
        \frac{D_{_F}A^\mu}{D\tau}=\frac{DA^\mu}{D\tau}-u^\mu (A_\nu a^\nu)+a^\mu (A_\nu u^\nu),
\end{equation}
where we have excluded the terms associated with spatial rotation i.e., $\epsilon_{\mu\nu\rho}A^\rho$.} 
In other words, the Fermi derivative of the vector field $A^\mu$ is defined as the difference between $A^\mu$ at $(x^\alpha+dx^\alpha)$ and its Fermi-Walker transported value from $x^\alpha$ to $(x^\alpha+dx^\alpha)$ along the curve $x^\mu=x^\mu(\tau)$.
Thus, the Fermi derivative of spin vector $s^\mu$ associated with the inertial guidance gyro vanishes along the worldline: 
\begin{equation}
    \label{spin_FW_I}
	\frac{D_{_F}s^{\mu}}{D\tau}=\frac{Ds^\mu}{D\tau}-u^{\mu} \left(a_\nu s^\nu\right)=0.
\end{equation}
Since the Fermi derivative of $u^\mu$ also vanishes, multiplying the equation above by $h^{\alpha}_{\mu}$ reveals that the spatial rate of change of spin is zero:  
\begin{equation}
    \frac{D_{_F}s^{\alpha}}{D\tau}=h^{\alpha}_{\mu}\frac{Ds^\mu}{D\tau}=0.
\end{equation}
Thus, the spin of an inertial guidance gyroscope follows
\begin{equation}
    \label{spin_FW_II}
    s^\mu u_\mu=0, \quad s^\mu s_\mu=1, \quad \frac{Ds^\mu}{D\tau}=u^\mu (s_\nu a^\nu).
\end{equation}

To compare the relative orientation of the inertial gyros carried by two infinitesimally separated observers requires transporting the spin of one of them along the separation vector that joins them.
Since a spatial (temporal) vector remains spatial (temporal) under Fermi-Walker transport, one is interested in finding the difference between $s^\mu$ at $(x^\alpha+\xi^\alpha\:\delta\xi)$ and its Fermi-Walker transported value from $x^\alpha$ to $(x^\alpha+\xi^\alpha\: \delta\xi)$, for example.\footnote{The parallel propagator applied to $A^\nu$, for example, parallel transports it from $x^\alpha$ to $(x^\alpha+dx^\alpha)$: $$(\delta^\mu_\nu-\Gamma^{\mu}_{\nu\sigma }dx^\sigma)A^\nu.$$
Similarly, the Fermi-walker propagator applied to $A^\nu$, Fermi-Walker transports it from $x^\alpha$ to $(x^\alpha+dx^\alpha)$:
$$\left[\delta^{\mu}_{\nu}-\left(\Gamma^{\mu}_{\nu\sigma}-u^{\mu}\nabla_{\sigma}u_{\nu}+u_{\nu}\nabla_{\sigma}u^{\mu}\right)dx^\sigma\right] A^\nu.$$} 
Thus, if the Fermi derivative of spin along the separation vector is zero, then these inertial gyros are aligned.\footnote{Since $s^\mu$ is Fermi-Walker transported along the worldline, one can use either separation vector or spatial separation to compare the relative orientation. From eq.(\ref{spin_FW_II}) and (\ref{spin_FW_III}), one can show that
\begin{equation}
	\frac{D_{_F}s^\mu}{D\xi_{_\perp}}=\frac{D_{_F}s^\mu}{D\xi}.
\end{equation}} 
Here, the Fermi derivative of $A^\mu$ along an arbitrary curve $x^\alpha=x^\alpha(\lambda)$ generalizes to
\begin{equation}
    \label{general_Fermi_derivative}
    \frac{D_{_F}A^\mu}{D\lambda}=\frac{DA^\mu}{D\lambda} + \left(u^{\nu}\nabla_{\sigma}u^{\mu}-u^{\mu}\nabla_{\sigma}u^{\nu}\right)A_\nu \frac{dx^\sigma}{d\lambda}.
\end{equation}
As can be seen ${D_{_F}u^{\mu}}/{D\lambda}=0$.
Multiply both sides of the equation above by $h^\alpha_\mu$ and take $A^\mu=s^\mu$
\begin{equation}
    \label{spin_FW_III}
    \frac{D_{_F}s^\alpha}{D\lambda}=h^\alpha_\mu \frac{Ds^\mu}{D\lambda}.
\end{equation}
Thus, Fermi derivative of spin reduces to taking the spatial projection of corresponding covariant derivative. 
Finally, the change in spin $s^\mu$ under the Fermi-Walker transport around a closed contour is given by
\begin{equation}
    \Delta_{_F}s^\mu=-\oint D_{_F} s^\mu=-\oint h^\mu_\nu\;Ds^\nu,
\end{equation}
that is, the spatially projected covariant change in spin around a closed contour. 

\section{The frame-drag field}
The frame-drag field as the temporal evolution of the relative orientation between infinitesimally separated inertial gyroscopes was derived independently in \cite{Nichols:2011pu}, \cite{Estabrook:1964zk} and Appendix of \cite{Sachs1960}. Here, we rederive it using Fermi-Walker transport around the closed contour \textbf{C}$\mathbf{_{2}}$ (formed by the four-velocity and spatial separation fields) and identify a previously overlooked term as the differential Thomas precession, a purely special-relativistic effect.
Basically, in the absence of a gravitational field, the observers carrying inertial gyroscopes notice relative precession if their relative velocity and accelerations are not parallel.
Moreover, we define the frame-drag field as a constraint: the covariant sum of relative velocities around the closed contour \textbf{C}$\mathbf{_4}$ (formed by spatial vector fields with a temporal ``closer" at the second order if the vorticity is non-zero) is given by the frame-drag field and the component of momentum density (of fluid) tangential to the contour. 
This definition in integral form is consistent with the following definition in differential form: the frame-drag field is the spatial curl of the velocity gradient $\nabla_\nu u_\mu$. 
The frame-drag field has no Newtonian analog and its value, as usual, depends upon the observer. For example, in Schwarzschild spacetime, the frame-drag field is absent for the congruence of radial timelike geodesics but is present for timelike geodesics with circular spatial trajectories.

\subsection{As the relative ``gravitational" precession between inertial gyroscopes}
The change in $s^\mu$ under Fermi-Walker transport around the closed contour \textbf{C}$\mathbf{_2}$ gives the change in relative orientation of inertial gyroscopes carried by observers $1$ and $2$ in proper time $\delta\tau$ (see Fig.\ref{Physical_setup_I} and section \ref{FW_transport}):
\begin{equation}
	\frac{\Delta_{_F} s^\mu}{\delta\tau\delta\xi}=h^{\mu}_{\nu} \frac{D}{D\xi_{_\perp}} \left(h^{\nu}_{\sigma}\frac{Ds^\sigma}{D\tau}\right) - h^{\mu}_{\nu} \frac{D}{D\tau} \left(h^{\nu}_{\sigma}\frac{Ds^\sigma}{D\xi_{_\perp}}\right)+h^{\mu}_{\nu} \left(\xi_{\alpha} a^{\alpha}\right) \frac{Ds^\nu}{D\tau}.
\end{equation}
Since the spin is Fermi-Walker transported along the worldline (\ref{spin_FW_II}), this simplifies to 
\begin{equation}
	\label{Spin_FW_transport}
	\frac{\Delta_{_F} s^\mu}{\delta\tau\delta\xi}=-h^{\mu}_{\nu} \frac{D}{D\tau} \left(\frac{Ds^\nu}{D\xi_{_\perp}}\right)+a^{\mu} \left(s_\sigma\frac{Du^\sigma}{D\xi_{_\perp}}\right).
\end{equation}
To introduce curvature in the equation above, use the Ricci identity (\ref{Ricci_identity}) and eq.(\ref{spin_FW_II}):
\begin{equation}
	\label{Spin_parallel_transport}
	R^{\delta}_{\;\;\nu\rho\sigma} h^{\mu}_{\delta}s^\nu u^\rho \xi^\sigma = h^{\mu}_{\nu} \frac{D}{D\tau} \left(\frac{Ds^\nu}{D\xi_{_\perp}}\right)- \left(s^\alpha a_\alpha\right)\frac{Du^\mu}{D\xi_{_\perp}}.
\end{equation}
From eq.(\ref{Spin_FW_transport}) and (\ref{Spin_parallel_transport}), it turns out that the relative precession depends upon both curvature and special relativistic terms:
\begin{align}
	\frac{\Delta_{_F} s^\mu}{\delta\tau\delta\xi}&=-R^{\delta}_{\;\;\nu\rho\sigma} h^{\mu}_{\delta}s^\nu u^\rho \xi^\sigma+a^{\mu} \left(s_\sigma\frac{Du^\sigma}{D\xi_{_\perp}}\right)-\left(s^\alpha a_\alpha\right)\frac{Du^\mu}{D\xi_{_\perp}} \\
	&=-R^{\delta}_{\;\;\nu\rho\sigma} h^{\mu}_{\delta}s^\nu u^\rho \xi^\sigma+\epsilon^{\mu\nu\rho}\: s_{\nu}\;\epsilon_{\rho\sigma\delta}\: a^{\sigma}\frac{Du^\delta}{D\xi_{_\perp}}\label{FD_precession_I}
\end{align}
where the spatial triple product between acceleration, relative velocity and spin has been simplified using the following relationship:
\begin{equation}
	\label{spatial_triple_product}
	\epsilon^{\mu\nu\sigma}\epsilon_{\alpha\beta\sigma}=h^{\mu}_{\alpha}\:h^{\nu}_{\beta}-h^{\mu}_{\beta}\:h^{\nu}_{\alpha}.
\end{equation}
To further introduce the frame-drag field, multiply eq.(\ref{H_Riemann}) by $\epsilon^{\alpha\beta\mu}s_{\beta} \xi^\nu$ and use (\ref{spatial_triple_product})
\begin{equation}
	\label{FD_precession_II}
	\epsilon^{\alpha\beta\mu}s_{\beta}\:\mathbb{H}_{\mu\nu}\xi^\nu=R^{\rho\sigma}_{\quad\nu\delta}\: h^{\alpha}_{\rho}\:s_\sigma \xi^\nu u^\delta.
\end{equation}
Thus, from eq.(\ref{FD_precession_I}) and (\ref{FD_precession_II})
\begin{align}
	\frac{\Delta_{_F} s^\mu}{\delta\tau}&=\epsilon^{\mu\nu\rho}s_\nu\:\mathbb{H}_{\rho\sigma}\:\xi^\sigma\delta\xi + \epsilon^{\mu\nu\rho}\: s_{\nu}\;\epsilon_{\rho\sigma\delta}\: a^{\sigma}\frac{Du^\delta}{D\xi_{_\perp}}\delta\xi. \\
	&=\epsilon^{\mu\rho\nu}s_\nu \left[-\mathbb{H}_{\rho\sigma} - \epsilon_{\rho\alpha\delta}\: a^{\alpha}\; \nabla_\sigma u^\delta\right]\xi^\sigma\delta\xi.
\end{align}
Thus, the precessional angular velocity is given by the terms in the square bracket times the spatial separation. 
In other words, according to the observer $1$, the spin of the gyroscope carried by observer $2$ precesses with angular velocity $-H_{\rho\sigma}\:\xi^\sigma\delta\xi$ due to the frame-drag field, with angular velocity $4\pi\epsilon_{\rho\sigma\delta}\:\xi^\sigma q^{\delta}\:\delta\xi$ (i.e., spatial cross product of separation and matter momentum density) due to $q_\mu$ and with angular velocity $\epsilon_{\rho\sigma\delta}\left(Du^\sigma/D\xi\right)a^{\delta}\:\delta\xi$ (i.e., spatial cross product of relative velocity and acceleration) due to the differential Thomas precession.

\subsubsection{Relative Thomas precession}
In this section, using a specific example, we show that the relative Thomas precession calculated using the familiar techniques of special relativity is equivalent to $\epsilon_{\mu\nu\rho} (Du^\nu/D\xi) a^{\rho}$. We start by briefly reviewing Thomas precession.

In special relativity, curvilinear motion of a particle can be analyzed using instantaneous inertial rest frames; rather than using a single non-inertial comoving frame, one assigns, at each instant, an inertial frame comoving with the particle.
For example, consider a particle moving with three-velocity $\vec{v}$ at time $t$ and with $\vec{v}+\delta\vec{v}\:$ at $t+\delta t$ in an inertial lab frame $K$.
Now, let the inertial frames $K'$ and $K''$ be the rest frames of the particle at times $t$ and $t+\delta t$, respectively.
In other words, $K'$ and $K''$ are moving with three-velocities $\vec{v}$ and $\vec{v}+\delta\vec{v}\:$ with respect to $K$.
If the coordinate axes of $K'$ and $K''$ have the same orientation as the lab frame $K$, then $K'$ and $K''$ are related to $K$ via a single Lorentz transformation. 
However, the coordinate axes of $K'$ and $K''$ will not have the same orientation if $\vec{v}$ and $\delta\vec{v}$ are non-parallel, since they are related by a spatial rotation and a Lorentz transformation \cite{Jackson, Moller}.
It turns out that the coordinate axes of $K'$ must be rotated by \footnote{Assume that the origins of these inertial reference frames coincide at time $t=t'=t''=0$. Then the coordinates in lab frame are related to the coordinates in $K'$ by the Lorentz transformation
\begin{equation}
	\vec{r'}=\vec{r}+\vec{v}\left[ \left(\frac{\gamma -1}{v^2} \right)\left(\vec{r}\cdot\vec{v}\right) -\gamma t\right], \quad t'=\gamma\left(t-\vec{v}\cdot\vec{r} \right).
\end{equation}
And the connection between the coordinates of $K''$ and $K$ is given by the Lorentz transformation
\begin{equation}
	\vec{r''}=\vec{r}+\vec{w}\left[ \left(\frac{\Gamma -1}{w^2} \right)\left(\vec{r}\cdot\vec{w}\right) -\Gamma t\right], \quad t''=\Gamma\left(t-\vec{w}\cdot\vec{r} \right),
\end{equation}
where
\begin{equation}
	\vec{w}=\vec{v}+\delta\vec{v} \quad \text{and} \quad\Gamma=\frac{1}{\sqrt{1-w^2}}.
\end{equation}
Using these equations, one can show that \cite{Jackson,Moller}
\begin{equation}
	\vec{r''}=\vec{r'}\: + \:\left(\vec{r'} \times \Delta\vec{\Omega}_T\right)\;-\Delta\vec{v}\; t' \:, \quad t''=t'-\left(\vec{r'}\cdot \Delta\vec{v}\right), \quad\text{where}\quad \Delta\vec{v}=\gamma\left[ \delta \vec{v} + (\gamma-1) \left(\frac{\vec{v}\cdot \delta\vec{v}}{v^2}\right)\vec{v}  \right],
\end{equation}
up to the first order in $\delta\vec{v}$. Here, both $\Gamma$ and $w^{-2}$ are needed up to the first order in $\delta\vec{v}$ to find the relation above:
\begin{equation}
	\Gamma=\gamma+\gamma^3 \left(\vec{v}\cdot\delta\vec{v}\right), \quad \frac{1}{w^2}=\frac{1}{v^2}-\frac{2 \left(\vec{v}\cdot\delta\vec{v}\right)}{v^4}.
\end{equation}
}
\begin{equation}
	\label{Thomas_angular_separation}
	\Delta\vec{\Omega}_T=\left(\gamma-1\right)\frac{\vec{v}\times{\delta\vec{v}}}{v^2} \quad\text{where}\quad \gamma=\frac{1}{\sqrt{1-v^2}},
\end{equation}
up to first order in $\delta\vec{v}$ to have the same orientation as $K''$ and $K$ . 
In other words, we have assigned to the particle local coordinate axes that undergo spatial rotation.
If the particle is spinning and ``point-like," with its spin aligned with one of the coordinate axes of $K$ at time $t$, then at time $t+\delta t$ they will be misaligned by the angular separation $\Delta\vec{\Omega}_T$ according to the particle.

If one chooses an inertial frame $K'''$ as the rest frame of the particle at $t+\delta t$, with the same orientation of coordinate axes as $K'$, then $K'$ and $K'''$ are related by a Lorentz transformation. 
Thus, we have assigned to the particle local `non-rotating' coordinate axes that undergo Fermi-Walker transport. 
Since two successive Lorentz transformations are equivalent to a Lorentz transformation followed by a spatial rotation, the coordinate axes of $K$ must be rotated by $-\Delta\vec{\Omega}_T$ to have the same orientation as $K'$ and $K'''$ \footnote{Here, up to the first order in $\delta\vec{v}$
\begin{equation}
	\vec{r'''}=\vec{r'}-\Delta\vec{v}\: t'=\left(\vec{r}-\vec{r}\times\Delta\vec{\Omega}_T\right) + \left(\vec{w}-\vec{w}\times\Delta\vec{\Omega}_T\right)\left[\left(\frac{\Gamma-1}{w^2}\right)\vec{r}\cdot\vec{w}-\Gamma t \right],
\end{equation}
i.e., a Lorentz transformation followed by a spatial rotation.
}.
Thus, according to the lab frame $K$, the local `non-rotating' axes carried by the particle (or its spin if it is a ``point-like" spinning particle) precess with angular velocity
\begin{equation}
	\label{Thomas_precession}
	\vec{\omega}_{_\text{Thomas}}=\lim_{\delta t \to 0}-\frac{\Delta\vec{\Omega}_T}{\delta t}=\left(\frac{\gamma-1}{v^2}\right)\:{\vec{a}}\times \vec{v}=\frac{\gamma^2}{\gamma+1}\left(\vec{a}\times\vec{v}\right),
\end{equation}
where $\vec{a}=d\vec{v}/dt$ is the three-acceleration of the particle.

Now, consider a congruence of timelike observers in Minkowski spacetime in Cartesian coordinates. 
One of the approaches to find the angular velocity of the relative Thomas precession between observers $1$ and $2$ in Fig.\ref{Physical_setup_I} (b) is as follows.
Assign an inertial frame $K$ to stationary observers, and instantaneous inertial rest frames $K^{'}_1$ and $K^{'''}_1$ to observer $1$, and $K^{'}_2$ and $K^{'''}_2$ to observer $2$.
Here, $K^{'}_1$ and $K^{'''}_1$ are the rest frames of observer $1$ at times $t$ and $t+\delta t$, for example.
Moreover, the orientation of coordinate axes from $K\to K^{'}_{1} \to K^{'''}_{1}$ does not change. 
In other words, the local tetrads carried by observers $1$ and $2$ are Fermi-Walker transported along their respective worldlines.  
The difference between coordinate transformations from $K^{'''}_1 \to K^{'''}_2$ and $K^{'}_1 \to K^{'}_2$ in terms of angular displacement in spatial rotations at the second order (i.e., $\delta\tau\delta\xi$) would give the angular velocity of relative precession between the inertial gyros carried by observers $1$ and $2$, whereas the difference in terms of velocity in Lorentz transformations would give relative acceleration.
This approach is equivalent to the change in spin under Fermi-Walker transport around the closed contour \textbf{C}$_{\mathbf{2}}$ considered in the previous section.

Here, we follow a less computational approach using the following example. 
Consider a congruence of rigidly rotating observers in Minkowski spacetime in cylindrical coordinates, $ds^2=-dt^2+d\rho^2+\rho^2 d\phi^2+dz^2$, with constant angular velocity $\Omega=d\phi/dt$ and carrying inertial guidance gyroscopes.
Here, the velocity and acceleration fields, and the non-zero Christoffel symbols are given by
\begin{align}
	u^\mu&=\gamma\;\delta^{\mu}_{0} + \gamma\Omega\;\delta^{\mu}_{2} \quad\text{where}\quad\gamma=\frac{dt}{d\tau}=\frac{1}{\sqrt{1-\rho^2\Omega^2}}, \label{specific_example} \\
	a^\mu&= \left(-\rho\gamma^2 \Omega^2\right) \;\delta^{\mu}_{1}, \\
	\Gamma^{2}_{12}&=\frac{1}{\rho},\quad \Gamma^{1}_{22}=-\rho.
\end{align}
This congruence is expansion-free and shear-free with vorticity 
\begin{equation}
	\omega^\mu = \left(\Omega \gamma^2\right)\; \delta^{\mu}_{3}.
\end{equation}
The spatial cross product between acceleration and velocity gradient turns out to be
\begin{equation}
	\label{intermediate_relative_Thomas}
	-\epsilon_{\mu\nu\rho}\:a^\nu \nabla_{\sigma} u^\rho =\left(\rho\gamma^4\:\Omega^3\right) \delta_{\mu}^{3}\:\delta_{\sigma}^{1}. 
\end{equation}
Let the neighboring observers $1$ and $2$ orbit the center at coordinate radius $\rho$ and $\rho+\delta\rho$ respectively.
Using eq.(\ref{intermediate_relative_Thomas}), one can find the angular velocity at which the spin of inertial gyroscope carried by observer $2$ precess according to the one carried by $1$:
\begin{equation}
	\label{Relative_Thomas_precession}
	-\epsilon_{\mu\nu\rho}\:a^\nu \xi^\sigma_{_\perp} \nabla_{\sigma} u^\rho =\left(\rho\gamma^4\:\Omega^3\right) \delta_{\mu}^{3}.
\end{equation}
According to the inertial observers stationary with respect to the metric, the gyroscope carried by observer $1$ precesses with angular velocity
\begin{equation}
	{\omega}_{_\text{Thomas}}=-\left(\gamma-1\right)\Omega,
\end{equation}
along the the $z$ axis, from eq.(\ref{Thomas_precession}) and (\ref{specific_example}).
Note that in this example, $\omega^\mu$ is orthogonal to both $a^\mu$ and $\xi_{_\perp}^\mu$, and ${\omega}_{_\text{Thomas}}$ points in $-z$ direction for all observers forming the congruence.
Therefore, according to these inertial observers, the differential precessional angular velocity between inertial gyroscopes carried by observers $1$ and $2$ is given by
\begin{equation}
	 \frac{d}{d\rho}\:{\omega}_{_\text{Thomas}}=-\rho\gamma^3\:\Omega^3.
\end{equation}
One needs to multiply this result by $-dt/d\tau=-\gamma$ to transform it to the rest frame of observer $1$, i.e., to account for the time dilation and opposite angular displacement. 
(Recall that the infinitesimal angular displacement in the rest frame is $\Delta\vec{\Omega}_T$ whereas in the lab frame is $-\Delta\vec{\Omega}_T$, and we divided $-\Delta\vec{\Omega}_T$ by the time duration $\delta t$ in the lab frame to get (\ref{Thomas_precession})).
As can be seen, this result, matches with (\ref{Relative_Thomas_precession}).

\subsection{As the covariant sum of relative velocities around a closed contour}
The change in $u^\mu$ under parallel transport around the closed contour \textbf{C}$\mathbf{_{4}}$ is given by
\begin{equation}
	\label{sum_of_relative_velocities}
	\frac{\Delta u^\mu}{\delta\xi\delta\zeta}=-R^{\mu}_{\;\;\nu\rho\sigma}\: u^{\nu} \xi^{\rho}_{_\perp} \zeta^{\sigma}_{_\perp}=\frac{D}{D\zeta_{_\perp}}\left(\frac{Du^\mu}{D\xi_{_\perp}}\right)-\frac{D}{D\xi_{_\perp}}\left(\frac{Du^\mu}{D\zeta_{_\perp}}\right)-2a^\mu \left(\omega_{\alpha\beta} \xi^{\alpha} \zeta^{\beta} \right).
\end{equation}
To introduce the frame-drag field in the equation above, use eq.(\ref{Weyl_decomposition}) and (\ref{EM_Weyl_II}) to get
\begin{equation}
	\label{energy_flux_frame_drag}
	C_{\alpha\beta\gamma\delta}\: h^{\alpha}_{\mu}\: u^\beta \: h^{\gamma}_{\nu}\: h^{\delta}_{\rho}=\epsilon_{\nu\rho\sigma}\: H^{\sigma}_{\mu}.
\end{equation}
From eq.(\ref{Weyl}), (\ref{H_Riemann}) and (\ref{energy_flux_frame_drag}), it can be shown that
\begin{equation}
	\label{Frame_drag_matter}
	R_{\alpha\beta\gamma\delta}\: h^{\alpha}_{\mu}\: u^\beta \: h^{\gamma}_{\nu}\: h^{\delta}_{\rho}=\epsilon_{\nu\rho\sigma}\:\mathbb{H}^{\sigma}_{\;\;\mu}.
\end{equation}
Thus, substituting eq.(\ref{congruenceII}) and (\ref{Frame_drag_matter}) into (\ref{sum_of_relative_velocities}) yields
\begin{equation}
	-\mathbb{H}^{\delta\mu}\:\epsilon_{\delta\rho\sigma}\xi^{\rho}\zeta^{\sigma}=\frac{D}{D\zeta_{_\perp}}\left(\frac{Du^\mu}{D\xi_{_\perp}}\right)-\frac{D}{D\xi_{_\perp}}\left(\frac{Du^\mu}{D\zeta_{_\perp}}\right)-2a^\mu \left(\epsilon_{\alpha\beta\delta}\:\omega^\delta \xi^{\alpha} \zeta^{\beta} \right).
\end{equation}
Now, take $\lambda^\mu$ as the spatial cross product between separation vectors
\begin{equation}
	\label{dual_to_C_4}
	\lambda_\mu = \epsilon_{\mu\nu\sigma} \xi^\nu \zeta^\sigma \implies \epsilon^{\mu\nu\sigma} \lambda_\sigma = \xi^{\mu}_{_\perp} \zeta^{\nu}_{_\perp}-\xi^{\nu}_{_\perp} \zeta^{\mu}_{_\perp},
\end{equation}
and substitute it in the equation above (or directly into (\ref{sum_of_relative_velocities})) to get
\begin{equation}
	\mathbb{H}^{\delta\mu}\lambda_{\delta}=h^{\mu}_{\nu}\frac{D}{D\xi_{_\perp}}\left(\frac{Du^\nu}{D\zeta_{_\perp}}\right)-h^{\mu}_{\nu}\frac{D}{D\zeta_{_\perp}}\left(\frac{Du^\nu}{D\xi_{_\perp}}\right)+2a^\mu \left(\omega_{\alpha} \lambda^{\alpha} \right).
\end{equation}
Thus, $\mathbb{H}_{\mu\nu}$ measures the failure of $u^\mu$ to be parallel transported around the contour \textbf{C}$\mathbf{_4}$.
The last term on the right side of the equation above represents the change in $u^\mu$ under parallel transport along the ``closer of the contour."
This term vanishes if $u^\mu$ is parallel transported along the temporal ``closer," i.e., when $a^\mu=0$.
It also vanishes if the ``closer" vanishes, i.e., when the vorticity is either zero or orthogonal to the separation vectors at worldpoint $A$, so one can construct a purely spatial contour.
From Fig.\ref{Physical_setup_II}, $\mathbb{H}_{\mu\nu}$ can be further defined as follows: the covariant sum of relative velocities around the closed contour \textbf{C}$\mathbf{_4}$ is given by $\mathbb{H}_{\mu\nu}\lambda^\mu$, where $\lambda^\mu$ is the spatial cross product at the world point $A$ between separation vectors forming the contour.  
In other words, $\mathbb{H}_{\mu\nu}$ captures the path dependence of relative velocities: the difference between the relative velocity of observer $3$ with respect to $1$ calculated via observer $4$ versus $2$ is given by $\mathbb{H}_{\mu\nu}\lambda^\mu$. 
When the vorticity and acceleration are non-zero, the ``closer" ensures that the difference in relative velocity via two different paths is calculated at the same proper time (i.e., at the worldpoint $C''$).

\subsubsection{As the spatial curl of the velocity gradient}
In (1+3) covariant decomposition scheme, the spatial curl of a vector field $A^\alpha$ is defined as
\begin{equation}
	\text{curl}\: A_{\mu}=\epsilon_{\mu\nu\rho}\; h^{\nu\alpha}\; h^{\rho}_{\beta}\; \nabla_{\alpha} A^{\beta}, 
\end{equation}
and of a rank-2 tensor $B^{\alpha\beta}$ is defined as \cite{Maartens:1997fg}
\begin{equation}
	\text{curl}\: B_{\mu\nu}=\epsilon_{\alpha\beta(\mu}\; h^{\alpha\rho}\; h^{\sigma}_{\nu)}\;h^{\beta}_{\delta}\; \nabla_{\rho} B^{\;\;\delta}_{\sigma}=\epsilon^{\rho}_{\;\;\delta(\mu}\; h^{\sigma}_{\nu)}\; \nabla_{\rho} B^{\;\;\delta}_{\sigma}.
\end{equation}
Thus, the spatial curl of four-velocity is the vorticity
\begin{equation}
	\text{curl}\: u^{\mu}=\epsilon^{\mu\nu\rho}\; h^{\alpha}_{\nu}\; h^{\beta}_{\rho}\; \nabla_{\alpha} u_{\beta}=\epsilon^{\mu\alpha\beta}\; \nabla_{\alpha} u_{\beta}=-2\omega^{\mu}
\end{equation}
and the spatial curl of the velocity gradient $\nabla_{\alpha}u_{\beta}$ is
\begin{equation}
	\text{curl}\: B^{\mu\nu}=\epsilon^{\rho\delta(\mu}\; h_{\sigma}^{\nu)}\; \nabla_{\rho} \nabla_{\delta}\: u^{\sigma}=\epsilon^{\rho\delta(\mu}\; u^{\nu)}\; u_{\sigma}\nabla_{\rho} \nabla_{\delta}\: u^{\sigma}+\epsilon^{\rho\delta(\mu}\; \nabla_{\rho} \nabla_{\delta}\: u^{\nu)}, 
\end{equation}
which, from the equations (\ref{Ricci_identity}) and (\ref{H_Riemann}), reduces to
\begin{equation}
	\frac{1}{2}\:\epsilon^{\rho\delta(\mu}\; u^{\nu)}\; u_{\sigma}\: R_{\rho\delta}^{\quad\sigma\gamma}\: u_{\gamma}+ \frac{1}{2}\:\epsilon^{\rho\delta(\mu}\; R_{\rho\delta}^{\quad\nu)\gamma} \: u_{\gamma}=\mathbb{H}^{(\mu\nu)}=H^{\mu\nu}.
\end{equation}
Thus, the frame-drag field can be defined as the spatial curl of velocity gradient, which is consistent with the previous section except for the symmetrization.

\section{The tidal field}
The tidal field, defined as the temporal evolution of the relative velocity between infinitesimally separated observers, is well known \cite{Misner:1973prb,Hawking}. 
Here, we define it as a constraint: the failure to align the local non-rotating spatial triads carried by observers forming the contour \textbf{C}$\mathbf{_4}$ (see Fig.\ref{Physical_setup_I}) is determined by the tidal field, the fluid variables (energy density and anisotropic pressure), and the spatial cross product of the relative velocities at the worldpoint $A$.
Although the tidal field has a Newtonian analog in terms of relative acceleration, this effect is absent and may provide a test of general relativity. 

\subsection{As relative ``gravitational" acceleration between observers}
The change in $u^{\mu}$ under parallel transport around the closed contour \textbf{C}$_\mathbf{2}$ (i.e., $AA'B''B$), from Figs. \ref{Physical_setup_I} and \ref{Physical_setup_II}, is given by
\begin{equation}
    \frac{\Delta u^{\mu}}{\delta\tau\delta\xi}=-R^{\mu}_{\;\;\nu\rho\sigma}u^{\nu}u^{\rho}\xi^{\sigma}=\frac{D}{D\xi_{_\perp}}\left(\frac{Du^{\mu}}{D\tau}\right)-\frac{D}{D\tau}\left(\frac{Du^{\mu}}{D\xi_{_\perp}}\right)+a^\mu \left(\xi_\alpha a^\alpha\right).
\end{equation}
Using eq.(\ref{E_Riemann}) this relationship can be rewritten as \cite{Hawking}
\begin{equation}
    \label{tidal_relative_acceleration}
    -\mathbb{E}^{\mu}_{\nu}\;\xi^{\nu}\:\delta\xi=h^{\mu}_{\nu}\: \frac{D}{D\tau} \left(\frac{Du^{\nu}}{D\xi_{_\perp}}\right)\:\delta\xi- \left[h^{\mu}_{\nu}\frac{Da^\nu}{D\xi_{_\perp}}+a^{\mu} \left(\xi_{\nu} a^{\nu}\right)\right]\delta\xi.
\end{equation}
The first term on the right is the total relative acceleration (see eq.(\ref{relative_acceleration})) whereas the terms in square bracket represent relative acceleration due to `non-gravitational' interactions.  
For example, in vacuum, if the observers $1$ and $2$ are in rockets, they can avoid relative acceleration by using thrusters to cancel the relative acceleration due to the tidal field.

Although the vector fields $u^\mu$ and $\xi^\mu_{_\perp}$ don't form a closed contour, the relative velocity between the two observers calculated using two different approaches (spatial rate of change of $\xi^\mu_{_\perp}$ versus change in $u^\mu$ along $\xi^\mu_{_\perp}$) is equivalent, i.e.,
\begin{equation} 
	    h^{\mu}_{\nu}\: \frac{D\xi^{\nu}_{_\perp}}{D\tau}=\frac{Du^\mu}{D\xi_{_\perp}},
\end{equation}
since the ``closer of the contour" is of the second order.
In the absence of the curvature, however, the spatial rate of change in relative velocity, in general, is not the same as the spatial change in $a^\mu$ along $\xi^{\nu}_{_\perp}$
\begin{equation} 
	h^{\mu}_{\nu}\: \frac{D}{D\tau} \left(\frac{Du^{\nu}}{D\xi_{_\perp}}\right)\neq h^{\mu}_{\nu}\frac{D}{D\xi_{_\perp}}\left(\frac{Du^\nu}{D\tau}\right).
\end{equation}   
In other words, the relative acceleration here, calculated using spatial change in $a^\mu$ along $\xi^{\nu}_{_\perp}$ must include the contribution due to the ``closer of the contour" and is given by the square bracket in eq.(\ref{tidal_relative_acceleration}).

\subsection{As the sum of relative orientations of inertial gyros around a closed contour}
The change in spin $s^\mu$ under Fermi-Walker transport around the closed contour \textbf{C}$_{\mathbf{4}}$ (i.e., $ABC''D$), from Figs.\ref{Physical_setup_I} and \ref{Physical_setup_II}, is given by
\begin{equation}
    \label{FW_closed_II}
    \frac{\Delta_{_F}s^\mu}{\delta\xi\:\delta\zeta}=\frac{D_{_F}}{D\zeta_{_\perp}} \left(\frac{D_{_F}s^\mu}{D\xi_{_\perp}}\right) - \frac{D_{_F}}{D\xi_{_\perp}} \left(\frac{D_{_F}s^\mu}{D\zeta_{_\perp}}\right)-2\omega_{\alpha\beta}\xi^\alpha\zeta^\beta \left(\frac{D_{_F}s^\mu}{D\tau} \right).
\end{equation}
Since the spin is Fermi-Walker transported along the worldlines, the ``closer of the contour" won't affect the result and from eq.(\ref{spin_FW_I}), (\ref{spin_FW_II}) and (\ref{spin_FW_III}), the equation above simplifies to
\begin{equation}
    \label{FW_closed_III}
    \frac{\Delta_{_F}s^\mu}{\delta\xi\:\delta\zeta}=h^\mu_\nu \frac{D}{D\zeta_{_\perp}} \left(h^\nu_\sigma\frac{Ds^\sigma}{D\xi_{_\perp}}\right) - h^\mu_\nu\frac{D}{D\xi_{_\perp}} \left(h^\nu_\sigma\frac{Ds^\sigma}{D\zeta_{_\perp}}\right).
\end{equation}
To introduce curvature in the equation above, use the Ricci identity (\ref{Ricci_identity}) and eq.(\ref{spin_FW_II}):
\begin{equation}
    \label{parallel_closed_II}
        R^{\alpha}_{\;\;\beta\gamma\delta}h^{\mu}_{\alpha}s^\beta \xi^\gamma_{_\perp} \zeta^\delta_{_\perp}= h^{\mu}_{\alpha}\frac{D}{D\xi_{_\perp}} \left(\frac{Ds^\alpha}{D\zeta_{_\perp}}\right) - h^{\mu}_{\alpha}\frac{D}{D\zeta_{_\perp}} \left(\frac{Ds^\alpha}{D\xi_{_\perp}}\right).
\end{equation}
From eq.(\ref{spin_FW_II}), (\ref{FW_closed_III}) and (\ref{parallel_closed_II})
\begin{align}
    \frac{\Delta_{_F} s^\mu}{\delta\xi\:\delta\zeta}&=-R^{\alpha}_{\;\;\beta\gamma \delta}\:h^{\mu}_{\alpha}\:s^\beta\xi^\gamma_{_\perp}\zeta^\delta_{_\perp}-\left(\frac{Du^\mu}{D\zeta_{_\perp}}\frac{Du^\alpha}{D\xi_{_\perp}}-\frac{Du^\alpha}{D\zeta_{_\perp}}\frac{Du^\mu}{D\xi_{_\perp}}\right)s_\alpha,  \label{Gauss-Codazzi} \\
    &=-R^{\alpha}_{\;\;\beta\gamma\delta}\:h^{\mu}_{\alpha}\:s^\beta\xi^\gamma_{_\perp}\zeta^\delta_{_\perp}-\epsilon^{\mu\alpha\beta} s_\alpha\; \epsilon_{\beta\sigma\delta}\frac{Du^\sigma}{D\zeta_{_\perp}}\frac{Du^\delta}{D\xi_{_\perp}}.     \label{Gauss-Codazzi_more}
\end{align}
To introduce the tidal field, first use eq.(\ref{Weyl_decomposition}) to get 
\begin{equation}
    C_{\alpha\beta\gamma\delta}h^{\alpha\mu}h^{\beta\nu}h^{\gamma}_{\rho}h^{\delta}_{\sigma}=4\:h^{\;\;[\mu}_{[\rho} E^{\;\;\nu]}_{\sigma]},
\end{equation}
and then use (\ref{Weyl}), (\ref{EM_decomposition}) and (\ref{fluid_components}) to get
\begin{equation}
	\label{Riemann_spatial_projection}
    R_{\alpha\beta\gamma\delta}h^{\alpha\mu}h^{\beta\nu}h^{\gamma}_{\rho}h^{\delta}_{\sigma}=4\:h^{\;\;[\mu}_{[\rho} \bigg( E^{\;\;\nu]}_{\sigma]} +4\pi\: \Pi^{\;\;\nu]}_{\sigma]} + \frac{4\pi\rho}{3}\: h^{\;\;\nu]}_{\sigma]}\bigg) \neq 4\:h^{\;\;[\mu}_{[\rho}\; \mathbb{E}^{\;\;\nu]}_{\sigma]},  
\end{equation}
in general. 
Finally, from eq.(\ref{spatial_triple_product}), (\ref{dual_to_C_4}) and (\ref{Riemann_spatial_projection}), we get
\begin{align}
	R^{\alpha}_{\;\;\beta\gamma\delta}h^{\mu}_{\alpha}s^\beta \xi^\gamma_{_\perp} \zeta^\delta_{_\perp}&=2\:h^{\;\;[\mu}_{\rho} \bigg( E^{\;\;\nu]}_{\sigma} +4\pi\: \Pi^{\;\;\nu]}_{\sigma} + \frac{4\pi\rho}{3}\: h^{\;\;\nu]}_{\sigma}\bigg)s_\nu\: \epsilon^{\rho\sigma\delta} \lambda_{\delta}, \\
	&=\epsilon^{\mu\alpha\beta}s_\alpha\: \epsilon_{\beta\gamma\nu} \Big(E^{\gamma}_{\rho}\: h^{\nu}_{\sigma} + 4\pi\: \Pi^{\gamma}_{\rho}\: h^{\nu}_{\sigma} \Big) \epsilon^{\rho\sigma\delta} \lambda_{\delta} + \frac{8\pi\rho}{3}\epsilon^{\mu\nu\delta}s_\nu\lambda_\delta, \\
	&=-\epsilon^{\mu\alpha\beta}s_\alpha\:\left(E_{\beta\delta} + 4\pi\: \Pi_{\beta\delta} - \frac{8\pi\rho}{3} h_{\beta\delta}\right) \lambda^\delta.
\end{align}
Substituting this into (\ref{Gauss-Codazzi_more}) yields
\begin{equation}
	\label{tidal_constraint}
	\frac{\Delta_{_F} s^\mu}{\delta\xi\:\delta\zeta}=-\epsilon^{\mu\alpha\beta}s_\beta\:\left[\left(E_{\alpha\delta} + 4\pi\: \Pi_{\alpha\delta} - \frac{8\pi\rho}{3} h_{\alpha\delta}\right) \lambda^\delta+\epsilon_{\alpha\sigma\delta}\:\frac{Du^\sigma}{D\xi_{_\perp}}\frac{Du^\delta}{D\zeta_{_\perp}} \right].
\end{equation}
Thus, in general, one cannot align the spins of inertial gyroscopes carried by the observers forming the closed contour \textbf{C}$\mathbf{_4}$. 
In other words, the local `non-rotating' spatial triads (or local inertial frames if the worldines are geodesics) carried by the four observers in Fig.\ref{Physical_setup_I} cannot be aligned or their relative orientations around \textbf{C}$\mathbf{_4}$ don't add up to zero in general. 

The last term on the right side of the equation above is the spatial cross-product between the relative velocities of observers $2$ and $4$ with respect to $1$ whereas $\lambda^\mu$ is the spatial cross-product between the corresponding separations. The former vanishes if the relative velocities are parallel and its effect vanishes if the spin is chosen to be orthogonal to both these relative velocities at the worldpoint $A$ \footnote{Consider a flat spacetime and apply successive Lorentz transformations around the loop \textbf{C}$\mathbf{_4}$ by assigning the instantaneous inertial rest frames to five worldpoints in Fig.\ref{Physical_setup_I} (c), for example.
From eq.(\ref{sum_of_relative_velocities}) and (\ref{tidal_constraint}), one can infer that the coordinate axes of instantaneous inertial rest frame at worldpoint $A$ and the same frame undergoing Lorentz transformations around the loop will not align. The angular separation between them, at the second order, is given by $\epsilon_{\alpha\sigma\delta}\:\left({Du^\sigma}/{D\xi_{_\perp}}\right) \left({Du^\delta}/{D\zeta_{_\perp}}\right)$. 
}. 

Here, the spin, under the Fermi-Walker transport around \textbf{C}$\mathbf{_4}$, is infinitesimally rotated (i.e., second order $\delta\xi\delta\zeta\:$) by $-E_{\alpha\delta}\lambda^\delta$ due to the tidal field, by $-4\pi\:\Pi_{\alpha\delta}\lambda^\delta$ due to the fluid anisotropic pressure, by $(8\pi\rho/3)\:\lambda_\alpha$ due to the fluid energy density and by the spatial cross-product of relative velocities. 
 Thus, if $s^\mu$ carried by observer $1$ in vacuum is orthogonal to relative velocities of observers $2$ and $4$ at a particular instance, it's failure to Fermi-Walker transport around the closed loop \textbf{C}$\mathbf{_4}$ is measured by the tidal field.

\subsubsection{Vanishing vorticity and the Gauss-Codazzi equation}
The Gauss-Codazzi equation tells how a three-dimensional hypersurface is embedded in a four-dimensional spacetime by relating the intrinsic and extrinsic curvatures of the former with the curvature of the latter \cite{Poisson,Ellis:1971pg}.

If the congruence is hypersurface orthogonal, i.e. $\omega_\mu=0$, then the spacelike hypersurface $\tau=$constant is purely spatial and one can construct a closed infinitesimal spatial contour. 
For this case, it turns out that the eq.(\ref{Gauss-Codazzi}) represents the Gauss-Codazzi equation where the change in spin under Fermi-Walker transport around a closed spatial contour is given by the three-dimensional curvature tensor intrinsic to the spatial hypersurface.
The second term on the right of (\ref{Gauss-Codazzi}) is related to the extrinsic curvature of hypersurface.

At a particular instant $\tau_{_0}$, choose $\xi^\mu,$ $\zeta^\mu$ and $\chi^\mu$ to be purely spatial and perpendicular to one another.
These vectors span the spatial hypersurface $\Sigma\; (\tau=\tau_{_0})$ and are the coordinate basis in the region where the congruence is non-intersecting
\begin{equation}
    \mathbf{e}^{\mu}_{1}=\frac{\partial x^\mu}{\partial y^1}=\xi^{\mu}, \quad \mathbf{e}^{\mu}_{2}=\zeta^{\mu}, \quad \mathbf{e}^{\mu}_{3}=\chi^{\mu}.
\end{equation}
A component of curvature tensor, for example, under this coordinate transformation (indicated by prime), is given by
\begin{equation}
    R_{\mu\nu\rho\sigma}\:\zeta^{\mu}\chi^{\nu}\zeta^{\rho}\xi^{\sigma} =R'_{\:2321},
\end{equation}
and a component of the extrinsic curvature or the second fundamental form $K_{ab}$ which calculates the change in the normal to $\Sigma$ along the tangent to $\Sigma$ (see e.g., Fig. $10.3$ of \cite{Wald:1984rg}), is given by
\begin{equation}
\xi^{\mu}\zeta^{\nu}\nabla_{\mu}u_{\nu}=\xi^{\mu}\zeta^{\nu}\nabla_{\nu}u_{\mu}=K_{12}.
\end{equation}
Choose $s^\mu$ to be parallel to $\chi^\mu$, i.e.,
\begin{equation}
    s^\mu=\frac{\chi^\mu}{\sqrt{g_{\alpha\beta} \chi^{\alpha} \chi^{\beta}}},
\end{equation}
then from eq.(\ref{Gauss-Codazzi})
\begin{equation}
    \label{GC-I}
    \zeta^{\mu}\:\frac{\Delta_{_F}s_{\mu}}{\delta\xi\delta\zeta}=\frac{R'_{2321}}{\sqrt{g'_{33}}}+\frac{1}{\sqrt{g'_{33}}}\left(K_{12}K_{23}-K_{22}K_{13} \right),    
\end{equation}
and
\begin{equation}
    \label{GC-II}
    \xi^{\mu}\:\frac{\Delta_{_F}s_{\mu}}{\delta\xi\delta\zeta}=\frac{R'_{1321}}{\sqrt{g'_{33}}}+\frac{1}{\sqrt{g'_{33}}}\left(K_{11}K_{23}-K_{21}K_{13} \right).    
\end{equation}
Compare these two with the following Gauss-Codazzi equation \cite{Poisson,Ellis:1971pg} 
\begin{equation}
	\label{GC_main_I}
    {}^{(3)}R_{abcd}=R'_{abcd}+\left(K_{ad}K_{bc}-K_{ac}K_{bd} \right)    
\end{equation}
where ${}^{(3)}R_{abcd}$ is the intrinsic curvature of the hypersurface.  
It is easy to see from (\ref{GC-I}),(\ref{GC-II}) and (\ref{GC_main_I}) that 
\begin{equation}
	\zeta^{\mu}\:\frac{\Delta_{_F}s_{\mu}}{\delta\xi\delta\zeta}=\frac{{}^{(3)}R_{2321}}{\sqrt{g'_{33}}} \quad \text{and} \quad \xi^{\mu}\:\frac{\Delta_{_F}s_{\mu}}{\delta\xi\delta\zeta}=\frac{{}^{(3)}R_{1321}}{\sqrt{g'_{33}}}.
\end{equation}
Thus, the failure to align the spins of inertial gyroscopes around a spatial loop is given by the intrinsic curvature of the spatial hypersurface.
In general, the change in a spatial vector under the Fermi-Walker transport around a spatial loop is given by the intrinsic curvature of the spatial hypersurface.

\section{Concluding remarks}
For a given congruence of timelike curves, consider the following four infinitesimal closed contours. 
\textbf{C}$\mathbf{_1}$ formed by spacelike (separation vector) and timelike (four-velocity) vector fields, which represents the temporal evolution of two observers. 
\textbf{C}$\mathbf{_2}$ formed by temporal (four-velocity) and spatial (spatial separation) vector fields with a temporal ``closer" of the contour proportional to four-acceleration.
\textbf{C}$\mathbf{_3}$ formed by two spacelike (separation) vector fields representing four observers located at the vertices of the ``quadrilateral" corresponding to this contour. 
\textbf{C}$\mathbf{_4}$ formed by two spatial (spatial separation) vector fields with a temporal ``closer" of the contour proportional to vorticity. 

For this congruence, $Du^\mu$ along the temporal direction gives the four-acceleration of the observer and along the spatial direction gives the relative velocity between two infinitesimally separated observers.
Therefore, $\oint Du^\mu$ along the closed contour \textbf{C}$\mathbf{_2}$ gives the relative acceleration between these observers due to curvature, i.e., the tidal field and fluid variables (energy density, isotropic and anisotropic pressures). 
Furthermore, $\oint Du^\mu$ along the closed contour \textbf{C}$\mathbf{_4}$ gives the covariant sum of relative velocities between four infinitesimally separated observers (accounting for different proper times via ``closer of the contour" if the vorticity is non-zero). 
Unlike Newtonian mechanics, this covariant sum is non-zero due to the curvature, i.e., the frame-drag field and momentum density of the fluid. 

Similarly, for the (unit spatial) spin vector field $s^\mu$ assigned to inertial guidance gyroscopes, $D_{_F}s^\mu$ along the temporal direction is zero, whereas $D_{_F}s^\mu$ along the spatial direction gives the relative orientation of inertial gyroscopes carried by two neighboring observers. 
Therefore, $\oint D_{_F}s^\mu=\oint h^\mu_\nu\: Ds^\nu$ along the closed contour \textbf{C}$\mathbf{_2}$ gives the rate of change of relative orientation of these gyroscopes. 
The angular velocity of this differential precession can be found by taking the spatial cross product with the spin, i.e., $\epsilon_{\mu\nu\sigma}s^\nu \oint D_{_F}s^\sigma$.
In other words, the local `non-rotating' spatial triads carried by two neighboring observers differentially precess due to both curvature (the frame-drag field and momentum density of the fluid) and relative Thomas precession (spatial cross product between relative velocity and acceleration).
Finally, $\oint D_{_F}s^\mu$ along the closed contour \textbf{C}$\mathbf{_4}$ gives the sum of the relative orientations of the gyroscopes carried by four observers forming the contour.   
This sum is non-zero due to both the curvature (the tidal field, energy density and anisotropic pressure of the fluid) and non-parallel relative velocities of neighboring observers.
This effect due to the tidal field is not present in Newtonian physics. 
If the vorticity is zero, $\oint D_{_F}s^\mu$ along \textbf{C}$\mathbf{_4}$ gives the intrinsic curvature of the spatial hypersurface, which is related to the spacetime curvature (tidal field, energy density and anisotropic pressure of the fluid) and the extrinsic curvature via the Gauss-Codazzi equation.

\section*{Acknowledgement}
The author acknowledges the support of the Campbell Grant (GR570834) and the Physics Summer Research Funds at Harvey Mudd College. 
The author thanks Professors Aaron Zimmerman (at Building Astronomy in Texas held at University of Texas, Dallas), Abhay Ashtekar, and Kip Thorne (at Charles W. Misner Memorial Symposium held at the University of Maryland, College Park and at the Bruce J. Nelson Distinguished Speaker Series at Harvey Mudd College) for stimulating discussions.


\end{document}